\documentclass[showpacs,amsmath,amssymb,twocolumn,aps,
        superscriptaddress]{revtex4-1}
\usepackage{marginnote}
 \usepackage{graphicx}
\usepackage{epsfig}              
 \usepackage{pstricks}
 \usepackage{pst-plot}
  \usepackage{psfrag}
 \usepackage{dcolumn}
 \usepackage{bm}

\usepackage{color}
\usepackage{cancel}
\usepackage{ulem}

\def\rr#1{\textcolor{black}{#1}}

 \def\mn#1{}
 \def\mg#1{}

\begin{document} 
\title{Semiclassical Approach to the Physics of 
Smooth Superlattice Potentials in Graphene}

\author{J\"urgen Dietel}
\affiliation{Institut f\"ur Theoretische Physik,
Freie Universit\"at Berlin, Arnimallee 14, D-14195 Berlin, Germany}
\author{Hagen Kleinert}
\affiliation{Institut f\"ur Theoretische Physik,
Freie Universit\"at Berlin, Arnimallee 14, D-14195 Berlin, Germany}
\affiliation{ICRANeT, Piazzale della Repubblica 1, 10 -65122, Pescara, Italy}
\date{Received \today}

\begin{abstract}
 Due to the chiral nature of the Dirac equation, \rr{governing the dynamics 
of electrons in graphene},
overlying  of an electrical superlattice (SL) 
can open new Dirac points  on the Fermi-surface of the energy spectrum. 
These lead to novel low-excitation physical phenomena.
A typical example for such a system is neutral graphene 
with a symmetrical unidirectional SL.  
We show here that \rr{in smooth SLs}, a semiclassical approximation
provides a good 
mathematical description for particles.
Due to the one-dimensional nature of the unidirectional 
potential, a wavefunction description leads to 
a generalized Bohr-Sommerfeld quantization condition for 
the energy eigenvalues. In order to pave the way for the application 
of semiclassical methods to two dimensional SLs in general, 
we compare these  energy eigenvalues 
with \rr{those obtained from numerical} calculations, and 
with the results from a semiclassical Gutzwiller trace formula
via the 
beam-splitting technique. Finally, we calculate 
ballistic conductivities in general point-symmetric 
unidirectional SLs with one electron and one hole region 
in the fundamental cell showing only Klein scattering 
of the semiclassical wavefunctions.     

\end{abstract}

\pacs{03.65.Sq, 72.80.Vp, 73.21.Cd, 73.22.Pr}

\maketitle

\section{Introduction}

Suspended graphene samples exhibit high \rr{electron} mobilities, where  
ballistic transport is seen for samples 
up to the micron length \cite{Morozov1,Du1, Bolotin1}.
Within the tight-binding approximation, the graphene system 
shows two inequivalent momentum energy knots in the Brillouin 
zone at low energies, located at momenta  
$ {\bf K} $  and $ {\bf K}' $. An effective low-energy description 
around these points is given by a massless Dirac equation.    
Electrons close to 
these points are related to each other 
by time-inversion symmetry \cite{Castro1}.
The  effective quasi-relativistic Hamiltonian is then given by   
\begin{equation} 
 H =\hbar 
v_F \left(\frac{1}{i} \sigma_x  \partial_{x} + 
\frac{1}{i} \sigma_y \partial_y\right) 
+ V({\bf r})  \label{10}   
\end{equation} 
for electrons near the $ {\bf K} $ point. Here $ v_F $ is 
the electron velocity in graphene and $ V({\bf r}) $ 
denotes an external potential. 

In the energy spectrum of the  associated Schr\"odinger 
equation, minigaps are opened by the application of an overlying 
superlattice potential. New bands arise
and 
two of them may touch each other at certain momenta, 
showing up in new Dirac points. These points are classified 
by their local similarity of the energy spectrum with the spectrum 
of the massless Dirac equation. Besides this, they show locally   
 \rr{a chiral behavior
in the pseudospin expectation value 
($\langle \sigma_x \rangle, \langle \sigma_y \rangle $) as a 
function of the Bloch momentum} \cite{Brey1}.      

This was first claimed theoretically  
by Park 
{\it et al.}  \cite{Park1,Park5},
and in fact new Dirac points were  found experimentally quite 
recently for graphene on a hexagonal 
boron nitride substrate \cite{Yankowitz1}.  
New Dirac points  are found at momenta $ G_m/2 $, where  $ G_m $ 
is a reciprocal lattice wavevector 
of the SL. Their energies are 
$ E_D = \pm \hbar v_F |G_m|/2 $ \cite{Park1}. Due to their nonzero energies,
these new Dirac points cannot be observed experimentally 
in low-energy excitation experiments on neutral graphene. 

Later on, additional new Dirac points were found in 
theoretical analyses, all located at zero energy. 
The calculations were  
done for graphene with a superimposed 
unidirectional electrical superlattice potential \cite{Brey1, Park2}. 
Actually, these new Dirac points had already appeared earlier in the literature 
within the framework of an unidirectional 
SL  on a nanotube 
\cite{Talyanskii1}. For the most simple representation of a 
unidirectional SL step-potential $ V(x)=V  \chi(x) $,  where 
 $ \chi(x)={\rm sg}[\sin(2 \pi x/d)]  $, 
\rr{and 
$ {\rm sg}[x] $ denotes the 
sign of $ x $}, the lowest energy band 
is shown in Fig.~1. The quantity $ d $ denotes the wavelength of the 
SL. 
The curves were obtained from a precise
numerical  
diagonalization.
The full energy   
spectrum of the lowest band energy   
shows a mirror symmetry  at the $ p_x $ and $ p_y $ axes. 

The energy spectrum close to the Dirac points  is given 
by \cite{Barbier1, Dietel2} 
 \begin{equation} 
\epsilon_s  = \! \!  s v_F  \tilde{\alpha}^2_0 
 \sqrt{p^2_x + 
|\hat{\Gamma}|^2 p_y^2}      
   \,  ,         \label{12} \\
\end{equation}
with $ \alpha_{0}  = \rr{([V/v_F]^2} -
      p_y^2)^{1/2} d/2 \hbar  $, 
$ \hat{\Gamma}=\sin[\alpha_{0}]e^{i \alpha_0}/\alpha_{0} $, 
$ \tilde{\alpha}_0=\alpha_0/\tilde{V}$ where $ \tilde{V}= 
V  d/2 \hbar v_F $. The  
Bloch momenta in $ x $-direction \rr{lie in the Brillouin zone
$ -\pi /d \le p_x/\hbar  \le \pi/d $}. The parameter  
$ s$ distinguishes \rr{the conduction band  ($s=1$) from the valence 
band ($s=-1$)}. 
From this we deduce that the Fermi-velocity 
$ {\bf v}= \hbar \partial \epsilon_s/\partial {\bf p} $ 
is in general  anisotropic at the central valley 
Dirac point  \cite{Park3}.  
The new Dirac points are located 
at momenta $ p_x=0 $ and $ p_y d/2 \hbar  = 
\pm (\tilde{V}^2- (\pi n)^2)^{1/2} $ with \rr{integer}
$ n \in \mathbb{N} $, where $ p_y $ takes real values.     
Furthermore,  for \rr{momenta} beyond the new Dirac points 
we obtain $ |p_y| \gg V/v_F $ \rr{the energy} $ \epsilon_s \sim s  v_F |p_y| $ \cite{Dietel2}. 
   
These new 
zero-energy Dirac points \rr{differ from the above mentioned points} of 
Park {\it et al.} \cite{Park1}, since they are  
not located at momenta with certain fractions of the reciprocal lattice. 
It is well known that these momenta are part of the 
region where SL minibands are formed. The Dirac points 
are then touching points of two minibands.   
In the case of the zero-energy Dirac points, due to the unidirectional SLs,  
the pristine Dirac cones are deformed 
strongly due to electrical potential 
such that the electron and valence bands touch.    

These new Dirac points are 
located at zero energy, and for that reason they possess a number 
of new interesting transport 
properties  \cite{Brey1, Burset1, Barbier2, Sun1, Park4, Dietel1, Dietel2}. 
By an application of an 
additional magnetic field,  
new Quantum-Hall plateaus are found \cite{Park2}. \rr{In disordered SLs,
there may also exist interesting
}
localization phenomena 
\cite{Zhu1}.    

\begin{figure}
\begin{center}
\includegraphics[clip,height=5.5cm,width=8.5cm]{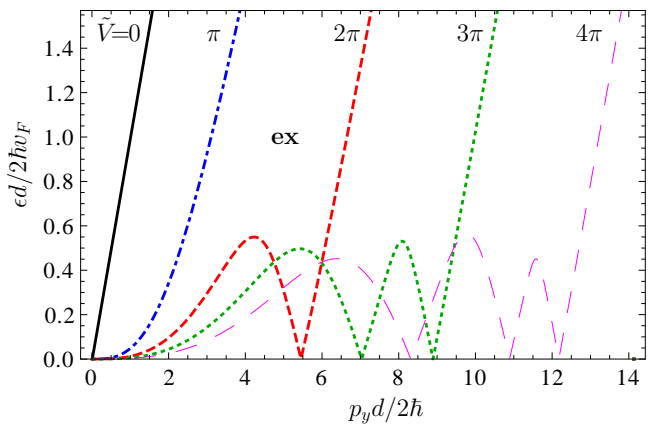}
\end{center}
\caption{Lowest energy band $ \epsilon d/2 \hbar v_F  $ 
for the Bloch momentum $ p_x =0 $ and several  
SL potentials $ V(x)=V\chi(x) $ as function of the dimensionless 
transversal momentum
$ p_yd/2 \hbar   $.}   
\end{figure}

A general understanding of the energy spectrum, and especially the 
location of new Dirac points, for general 
two dimensional non-unidirectional potentials is still missing.
Most interesting is the low-energy sector  of the energy spectrum
in neutral graphene. By  
taking into account that the zero-energy  
Dirac points show up only at large SL potentials,  
semiclassical methods may be used to determine 
the lowest energy bands for general smooth SLs. 
To justify this claim   we demand that
for unidirectional potentials the semiclassical condition
\begin{equation}    
\hbar v_F \frac{|(E-V) (V')|}{\sqrt{(E-V)^2 -(v_F p_y)^2}^3}  \ll 1 
\label{13} 
\end{equation} 
should be fulfilled, except at a few penetration points 
where 
\begin{equation} 
\sqrt{[E-V(x_p)]^2 -(v_F p_y)^2}=0.          \label{15} 
\end{equation}
Here $ V'(x) $ is the derivate of the potential $ V $ with respect to $x $.   
In classical mechanics, these points correspond to turning points. 
Due to the chiral nature of (\ref{10}) however this is no longer true. 
For example at the transverse momentum 
$ p_y = 0 $, the transmission probability 
at the penetration points is unity. Here the electron transform\rr{s} 
from a particle state (electron-like) to a hole state (positron-like) 
or vice versa.  
This is the analog of Klein's paradox \cite{Klein1, Hansen1, Greiner1} 
in relativistic quantum mechanics. \rr{In this context it was shown later} by 
Sauter \cite{Sauter1},
that the transmission probability is decreasing for 
non-zero momenta $ p_y $, 
and approaching  zero for  $ v_F p^2_y/ \hbar V'(x_p) \gg 1 $.

In order to prepare the semiclassical approach for smooth SLs,
in Sect.~II we shall
first \rr{discuss} the semiclassical wavefunctions 
of the problem. Then we derive 
transmission and reflection coefficients for 
electrons or holes, carrying out Klein's scattering analysis
through classically forbidden regions between 
two penetration points. 
We apply our results to the \rr{simplest }
unidirectional SL with one electron and one hole region in the fundamental cell,
showing only Klein scattering. Our  semiclassic results   
 for the lowest energy bands compare well with those obtained 
numerically. Furthermore,  we address 
the question whether one can construct SLs
within the semiclassical approximation which show the ability to 
focus electron beams. In order to see in which way a semiclassical 
approach could work also beyond the unidirectional SL case, 
in Sect.~III we consider 
the generalization of the Gutzwiller trace formula 
that includes also the beam splitting phenomenon
\rr{in the 
calculation of the 
semiclassical density of states}. \rr{When   
calculated from small-length orbits, the density of states
will permit us} to reconstruct the lowest energy band.   
Finally  in Sect.~IV, we shall calculate semiclassical 
conductivity formulas for smooth unidirectional SLs  
and compare our results with existing  
calculations in the literature. We restrict ourselves thereby to 
ballistic transport. Sect.~V and Sect.~VI contain a summary of the results.

\section{The semiclassical energy spectrum of the lowest-band} 

In the following we formulate the semiclassical approach  
to the
  quasi-relativistic 
Dirac equation
of electrons  in a unidirectional SL (\ref{10}). 
\rr{The energy 
spectrum will show mirror symmetry with respect to the transversal momentum 
$ p_y $ at the $ p_y=0 $ axis}.  
In the first subsection, the semiclassical 
solution of the eigenvalue problem will be obtained via
 the 
Bohr-Sommerfeld quantization condition for non-relativistic electrons.
We obtain  very good agreement for the energy spectrum of the lowest band  
with numerical results for 
the SL-deformed sinus potential.
In the second subsection, we consider an interesting counter example 
where the semiclassical approach fails.

\subsection{Generalized Bohr-Sommerfeld formalism} 
Starting point is the solution for the semiclassical wavefunction 
for the Hamiltonian (\ref{10}). 
This was previously done in the case of the relativistic 
Dirac-equation in Refs.~\cite{Rubinow1,Keppeler1}. 
We use the semiclassical Ansatz 
$ \rr{{\bf \Psi}_s}(x)= \sum_{n} \hbar^n {\bf \Psi}_n e^{i S(x)/ \hbar}$ in which 
$ {\bf \Psi}_n$, $S(x) $ are independent of $ \hbar $.  
Up to the order 
$ \hbar^0 $ we obtain 
\begin{equation} 
{\bf \Psi}_s(x) \! = \! \frac{|E-V(x)|}{v_F \sqrt{|p_x|}}
  \binom{ s  \frac{p_x -i p_y}{\sqrt{p_x^2 +p^2_y}}
}{1}   e^{i S(x)/\hbar + i \phi(x)}   e^{i p_y y/\hbar}  \label{20}  
\end{equation} 
with 
\begin{eqnarray} 
p_x(x) & = & \pm \sqrt{\frac{(E-V(x))^2}{v_F^2}-p_y^2}\,, \label{25}  \\ 
S(x) & = &  \int^x_{x_p} dx' p_x(x') \,,   \label{30}  \\ 
\phi(x) & = & - \frac{p_y}{2} \int^x_{x_p} dx' \frac{1}{p_x(x')} 
 \frac{ \partial_{x'} (E-V(x'))}{(E-V(x'))}    \,.     \label{40}  
\end{eqnarray} 
Here $ s= {\rm sg}[E-V(x)]  $, and  $ S(x) $ is the classical eikonal action of the particle. 
For neutral graphene, 
the states with $ s=1 $ are 
particle-like , and those with $ s=-1 $ are hole-like.  
The symbols
$ p_x $, $p_y $ denote
the momenta of the particle or hole in 
$ x $, $y $-direction. 
Note that for $ \phi= 0 $ and neglecting the vector part 
$ (s ( p_x -i p_y)/\sqrt{p_x^2 +p^2_y},1)^T $ in (\ref{20}), 
the wave function $ {\bf \Psi}_s(x) $ is the semiclassical solution of the 
massless quasi-relativistic Klein-Gordon wave equation. 
This means that $ \phi(x)  $ is a phase 
correction factor due to the chiral nature of the quasi-relativistic 
Dirac equation (\ref{10}). This phase factor has, of course, direct 
consequences on the semiclassical Bohr-Sommerfeld quantization condition 
\cite{Littlejohn1, Keppeler1}. 
Without proof we note that by taking into 
account also a homogeneous magnetic field this factor exactly cancels 
the Maslov index \cite{Maslov1} 
of the turning points such that the Landau level energy 
ladder starts at zero energy \cite{Carmier1}.   

We consider in this paper the simplest case
\rr{of small energies $ |E| \ll {\rm max}[|V(x)|] $  
where the 
scattering in a smooth  SL 
is 
mainly based on the so-called 
Klein tunneling for $ p_y^2 \lesssim {\rm max}[|V(x)|]/v_F $}. 
We shall discuss the situation in general also for 
larger $ p_y $ at the end of Sect.~IIA. 
 
\rr{We assume in the following that in the scattering process,
} incident particles are coming from  the left side of the 
tunnel region with a positive velocity and energy $ E-V(x)>0 $. 
The  energy $ E $ is conserved during the  
scattering processes considered in this paper. 
\rr{The particle then tunnels  
from the left penetration point $ x_{pL} $ (see Eq.(\ref{15}))  
into  a   classical forbidden region
 between the penetration points $ x_{pL} $ and $ x_{pR} $,
where $ \sqrt{[E-V(x)]^2 -(v_F p_y)^2} $ is imaginary}.
\rr{Beyond the penetration point $ x_{pR} $
on the right hand side of the tunnel region, 
it will reach the classical allowed hole region, where 
$ \sqrt{[E-V(x)]^2 -(v_F p_y)^2} $ is again real but now we have $ E-V(x)<0 $}.
In general in a Klein tunnel process a particle (hole) tunnel 
through a classical forbidden region into a hole (particle) region.  

\rr{Besides Klein tunneling, there are also 
conventional tunneling processes, for example a particle (hole) tunneling
through the full hole (particle) region at imaginary 
$ \sqrt{[E-V(x)]^2 -(v_F p_y)^2} $, i.e., where   
the particle (hole) does not change its signature $s$}.
We note that Klein tunneling is also referred to as interband scattering in the 
literature whereas conventional scattering is an innerband scattering event. 
One can find further discussions on the nature of scattering in graphene  
e.g. in Refs.~\onlinecite{Katsnelson2, Beenakker1,Lewkowicz1}.

By comparing (\ref{20})--(\ref{40})  with the definition 
of the penetration points (\ref{15}), we identify a singular behavior 
of the semiclassical wavefunction at 
these points. This means that there is still the freedom to linearly combine  
the basis of semiclassical wavefunction solutions in (\ref{20}) 
consisting of \rr{left- and right-moving} particles or holes 
with some yet 
undetermined numerical prefactors 
in every nonsingular potential sector. This freedom has to be fixed by further 
physical arguments. As in the quasi non-relativistic case, we achieve this    
by matching the wavefunction (\ref{20}) to the $ x \rightarrow \pm \infty$ 
asymptotics of the exact solution of (\ref{10}) 
for the linear potential $ V(x)= P (x-\overline{x}_p) $ where 
$P \approx \partial_{\overline{x}_p} V(\overline{x}_p)\equiv  V'(\overline{x}_p) $, with
$\overline{x}_{p} \equiv (x_{pR}+x_{pL})/2 $. 
\rr{In the following we assume that the SL 
is smooth in the classically forbidden region, meaning that 
$ V'(x)= P $ changes little  between the 
left and right penetration points $ x_{pL} $ and $ x_{pR} $ where $ P>0 $}. 
In order to solve this linear potential scattering 
problem we first use the Ansatz 
$ {\bf \Psi}(x) =\hbar v_F [\frac{1}{i} \sigma_x  \partial_{x} + \sigma_y p_y/\hbar 
-P x](1,1)^T \varphi(x) $. This leads to a Klein-Gordon-like 
differential equation for the wavefunction $ \varphi(x) $ in (\ref{10}),
in rescaled coordinates reading  
\begin{equation} 
\left[\partial_{x'}^2+ \frac{1}{4} x'^2 -\left(\frac{i}{2}  + \tilde{p}_y^2
\right)
\right] \varphi(x')= 0  \label{45} 
\end{equation}  
where the dimensionless transversal momentum square $  \tilde{p}_y^2 $ 
is given by
\begin{equation}  
\tilde{p}^2_y  =   \frac{v_F p_y^2 }{2 \hbar |P|}       \label{48} 
\end{equation} 
and \rr{$ x'\equiv(2 P/\hbar v_F)^{1/2} x $}. 
It can be solved 
with the help of special functions \cite{Abramowitz1}.  
By comparing the asymptotics of this solution for $ x \to \pm \infty $ 
with the semiclassical wavefunction (\ref{20}),
we obtain the transmission and reflection 
coefficients (\ref{50}), (\ref{60}) of the scattering 
\rr{of an 
electron incident from 
the left at the potential $ V(x)= P (x-\overline{x}_p) $}.
 After a lengthy
but straightforward calculation, the reflection 
$ \overrightarrow{R}_{\rm eh} $ and 
transmission $ \overrightarrow{T}_{\rm eh} $ coefficients are  found as 
\begin{eqnarray} 
 \rr{ \overrightarrow{T}_{\rm eh}} & = & e^{-i \pi {\rm sg}[p_y]/2} e^{-\pi \tilde{p}^2_y},  \label{50} \\
  \overrightarrow{R}_{\rm eh} & = & e^{i \vartheta(\tilde{p}^2_y)} 
\sqrt{1-e^{-2 \pi \tilde{p}^2_y}} \,,  
\label{60} 
\end{eqnarray} 
with 
\begin{equation} 
 \vartheta(\tilde{p}^2_y) =  - \pi/4 + {\rm arg}[\Gamma(i \tilde{p}_y^2)] 
+\tilde{p}_y^2- \tilde{p}_y^2 \ln(\tilde{p}_y^2) \,.
 \label{80} 
\end{equation}
Here $ \Gamma $ is the Gamma function.
In the reflection and transmission coefficients the arrow on top 
of the coefficients denote the direction of scattering, i.e. from left to 
right or vice versa. The suffixes $eh$ ($he$) denotes the case  
where on the left (right) hand side of the scattering region 
the electron is particle-like and on the right (left) hand side  
hole like.

Note that a similar calculation for smooth graphene np and npn 
junctions was carried out in Refs.~\cite{Cheianov1, Sonin1, Tudorovskiy1}. 
In Fig.~2, we show in Fig.~2 the functions $ |T| \equiv 
|\overrightarrow{T}_{\rm eh}(\tilde{p}_y^2)| $, $ |R| \equiv 
|\overrightarrow{R}_{\rm eh}(\tilde{p}_y^2)| $ and $ 3\pi/4 + 
\vartheta(\tilde{p}^2_y) $,  
where  $ \vartheta $ is an increasing function 
of $ \tilde{p}_y^2 $ with limiting values $ \vartheta(0)=-3 \pi/4 $ and 
$ \lim_{\tilde{p}_y^2 \to \infty}  \vartheta(\tilde{p}_y^2)=-\pi/2 $.  

\begin{figure}
\begin{center}
\includegraphics[clip,height=5.5cm,width=8.5cm]{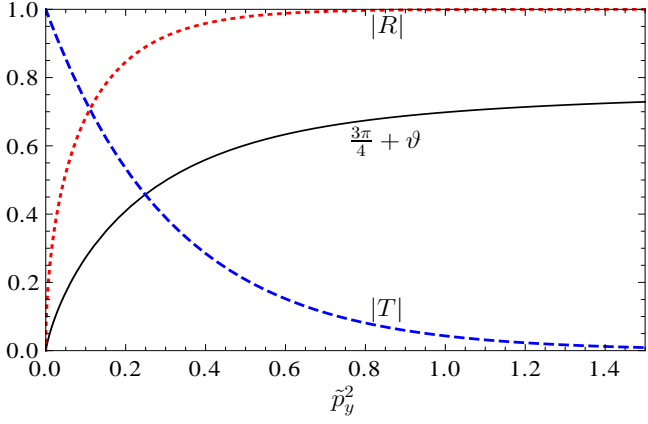}
\end{center}
\vspace{-2em}
\caption{Transmission function 
 $  |T| $, reflection function $ |R| $, and  
$ 3 \pi/4 +\vartheta $ as a function of $ \tilde{p}^2_y$.
}   
\end{figure}

The transmission and reflection coefficient 
for a hole incident from the right, but again   
$ V'(\overline{x}_p)>0  $, can then be determined from 
(\ref{50}) and (\ref{60}), using the invariance of (\ref{10}) 
under the transformation 
$ {\bf \Psi}(x) \rightarrow \sigma_z {\bf \Psi}^*(x) $ for fixed $ p_y $.
This leads to $   \overleftarrow{T}_{\rm eh} = \overrightarrow{T}_{\rm eh} $ and 
$   \overleftarrow{R}_{\rm eh} = \overrightarrow{R}^*_{\rm eh} $. 
The transmission and reflection coefficients for a potential with 
$ V'(x_p) \le 0 $ can be read off from the  above  coefficients
by using the substitution $ p_y \rightarrow -p_y $ in the 
corresponding expressions. This leads to 
$ \overrightarrow{T}_{\rm he}(p_y) = 
\overleftarrow{T}_{\rm eh}(-p_y) $, $ \overleftarrow{T}_{\rm he}(p_y)= 
\overrightarrow{T}_{\rm eh}(-p_y) $ and $ \overrightarrow{R}_{\rm he}(p_y) = 
\overleftarrow{T}_{\rm eh}(-p_y) $, $ \overleftarrow{R}_{\rm he}(p_y)= 
\overrightarrow{R}_{\rm eh}(-p_y) $.

For $ \hbar \to 0 $ or $ p_y \to \infty $, finally, 
we obtain from (\ref{50}), (\ref{60}) and (\ref{80}) that 
$  \overrightarrow{T}_{\rm eh} \to 0 $ and 
$ \overrightarrow{R}_{\rm eh}
= e^{-i \pi/2} $. Thus in this limit, we obtain the same 
reflection and transmission coefficient as  
for the reflection of a non-relativistic 
particle at a smooth potential barrier \cite{Kleinert1}.   


The matching procedure used above for the 
determination of the reflection and transmission coefficients requires that  
the potential changes little between the penetration points 
in the classically forbidden 
region. This has to be fulfilled even in the vicinity 
of the penetration points. We find from Eq.~(\ref{13}) that 
$V(x)$ should be almost constant where 
\begin{equation}
   \label{85}
\quad  |x - \overline{x}_p|
\lesssim \frac{x_{pR}-x_{pL}}{2}
+\sqrt{ \frac{\hbar v_F}{P}}  {\rm min}\left[1, \frac{1}{\tilde{p}_y^{1/3}}
\right]\,.  
\end{equation}
It is clear that a step-like SL does not fulfil this condition.  
Note that in the subsequent numerical calculations, we shall use 
$ P\rr{\approx} |V(x_{pR})-V(x_{pL})|/(x_{pR}-x_{pL}) $ for fixing  
$ \tilde{p}_y^2 $.

Next we calculate the energy spectrum for an unidirectional superlattice 
potential with two penetration points in the fundamental cell. This  
configuration is shown in Fig.~3. 
In order to calculate the eigenvalue spectrum,   
we have used a transfer matrix method. Note for example that 
the transfer matrix across the \rr{$i$}-th  
penetration point is  
\begin{equation} 
\underline{M}_i= \left(\begin{array} {cc} 
\overrightarrow{T}_i-\frac{\overleftarrow{R}_i\overrightarrow{R}_i}
{\overleftarrow{T}_i} & \frac{\overleftarrow{R}_i}
{\overleftarrow{T}_i} \\ -\frac{\overrightarrow{R}_i}
{\overleftarrow{T}_i} & \frac{1}
{\overleftarrow{T}_i}
\end{array} \right) \,.    \label{90} 
\end{equation} 
With the definitions 
\begin{equation} 
S_i(x)= (-1)^{i+1} 
\int_{x_{p_{i1}}}^{x}dx' \sqrt{\frac{(E-V(x'))^2}{v^2_F}-p_y^2}  
\label{100} ,
\end{equation} 
and 
\begin{equation} 
\underline{N}_i(x)= \left(\begin{array} {cc} 
e^{i \frac{S_i}{\hbar}(x)}  & 0        \\
0 & e^{- i \frac{S_i}{\hbar}(x)}
\end{array} \right)  ,   \label{110} 
\end{equation} 
the energy spectrum is given by 
$ {\rm det}[\underline{A}-
e^{ip_xd/\hbar}\underline{E}]= 0 $ where $ \underline{A}=  
\underline{M}_2\underline{N}_2(x_{p22})
\underline{M}_1\underline{N}_1(x_{p12})$, and 
$ \underline{E} $ denotes the unit matrix.   
The various intersection points $ x_{p_{ij}} $ are 
illustrated for a  sinus potential and $ E=0 $ in Fig.~3.  
From (\ref{40}), we obtain  
that the phase factors $ \phi(x)  $ in the  
transfer matrix $ \underline{N}_i $ are cancelled. By using \rr{once more the arguments}
following (\ref{80}) we obtain 
\begin{align} 
& \cos\left(\! \! \frac{S_1 \! + \! S_2}{\hbar} \! \! \right) \! - \! 
|R_1| |R_2|  
\cos \! \left(\!\!  \frac{S_1- S_2}{\hbar}\!   + 
\! {\rm arg}[\! \overrightarrow{R}_1\! ]
\! - \! {\rm arg}[\! \overrightarrow{R}_2 \! ]
\! \right) \nonumber \\
& 
\quad \quad =|T_1| |T_2| 
\cos\left( \frac{p_x d}{\hbar} \right)   ,
\label{120}     
\end{align}
where $ S_i \equiv  S_i(x_{pi2}) $. 
In order to obtain 
a particle-hole symmetry in the spectrum which is the requirement to find new 
Dirac points for $ E=0 $, we demand point symmetry 
of the SL-potential, i.e.  $ S_1=-S_2 $ for $ E=0 $.   
By using the fact that $ (1-|R_1| 
|R_2|)/|T_1| |T_2| > 1 $ 
for $ |R_1| \not= |R_2| $, we obtain that  
Eq.~(\ref{120}) can not be fulfilled for $ E = 0$
in the case of a potential which has no additional mirror symmetry with 
respect to the transversal momentum $ p_y $. This means that 
semiclassically we do not find any additional Dirac-point except  
the one for pristine graphene at $ p_x=p_y=0 $ for asymmetric potentials
where $\overrightarrow{R}_1 \not= -\overrightarrow{R}_2 $.
Numerically, this is seen using the  deformed sinus-potential 
\begin{equation} 
V(x)= V 
\sin\left[\frac{2 \pi (x-d/2)}{ d} \frac{(1+ a (x-d/2)^2)}{1+ a (d/2)^2}\right]
\label{130},
\end{equation}   
defined for $ 0 \le x \le d $, for calculating the lowest energy band  
by a numerical diagonalization, and compare the  
results with the semiclassical ones (\ref{120}).
In Fig.~3 we show these results for various deformed sinus potentials.

\begin{figure}
\begin{center}
\includegraphics[clip,height=5.5cm,width=8.5cm]{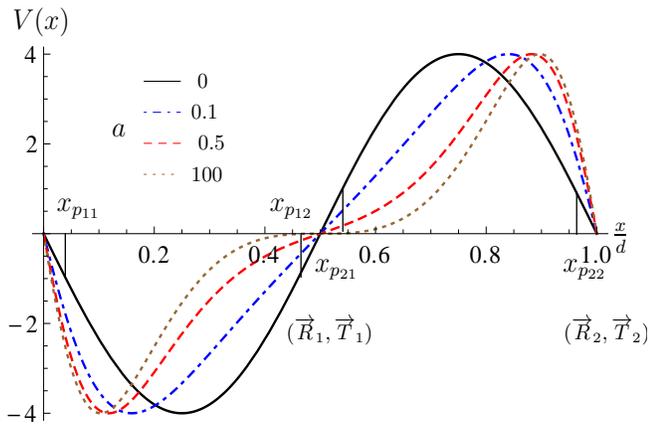}
\end{center}
\hspace*{0.2cm}
\vspace{-2em}
\caption{We show the deformed sinus potential (\ref{130}) for 
potenial strength $ V = 4   $ and deformation parameters 
$ a=0, 0.1,0.5, 100 $. The various penetration points 
are shown for the non-deformed sinus potential $ a=0 $ in the case that 
$ E =0 $ and  $ p_y v_F/ V  = 1 $.}  
\end{figure}

\begin{figure}
\begin{center}
\includegraphics[clip,height=4.5cm,width=8.5cm]{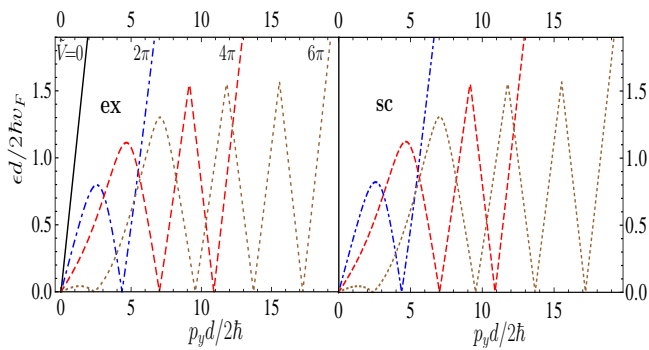}
\end{center}
\vspace{-2em}\caption{\rr{Lowest energy band at $ p_x=0 $ for 
the 
non-deformed sinus-potential
(\ref{130}), with $ a=0 $ and $ \tilde{V} =0, 2 \pi, 4 \pi, 6 \pi $. 
The left panel shows the numerical,
the right panel the semiclassical results 
(\ref{120}).} }  
\end{figure}

\begin{figure}
\begin{center}
\includegraphics[clip,height=4.5cm,width=8.5cm]{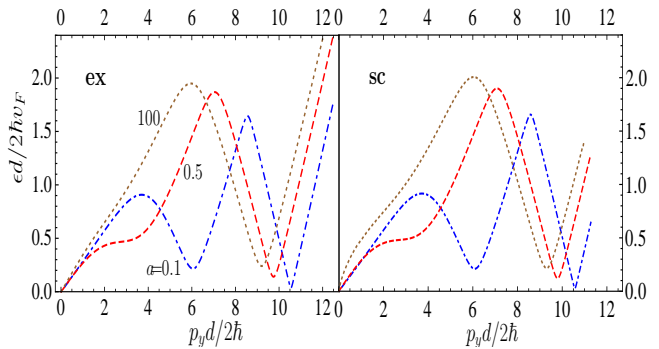}
\end{center}
\vspace{-2em}\caption{\rr{Lowest energy band at $ p_x=0 $ for
the 
deformed sinus-potential
(\ref{130}) with $ \tilde{V} =4 \pi $ 
and $ a=0.1 , 0.5, 100 $. 
The left panel shows the numerical,
the right panel the semiclassical results 
(\ref{120}).}}  
\end{figure}

In the left panel of Fig.~4,  we show  the lowest energy band for $ p_x=0 $, 
using the exact numerical diagonalization method for the various sinus 
potentials (\ref{130}), whereas   
in the right panel the corresponding semiclassical 
result (\ref{120}) is shown. The same is shown in Fig.~5 for various deformed 
sinus-potentials. 
The plots are characterized by
$ \tilde{V}(x) \equiv V(x)d/2 \hbar v_F $ 
since the energy spectrum (up to a simple rescaling of momentum and energy) 
as well as the conductivities depend mainly on this 
dimensionless potential.  In  both figures we obtain an almost perfect 
agreement between numerical diagonalization results and  
semiclassical lowest energy band.

As 
was already discussed following Eq.~(\ref{40}),
the Klein-scattering process dominantes over other  scattering processes 
for small energies $ E $ and 
momenta $ p_y $   for smooth SLs.
Let us elaborate this point further.      
\rr{First we can show using semiclassical methods similar to 
those applied to
the step-like case in Sect.~I,
that   
$ |\epsilon_s| \gtrsim  v_F |p_y| $ 
at large momenta where $ |p_y v_F| \gg {\rm max}[|V(x)|] $.}
On the other hand for small momenta 
where $ |p_y v_F| $ is smaller than the absolute value of possible 
local minima (maxima) of $ V(x)$ in the case of particles (holes), 
we obtain  that mainly Klein scattering 
\rr{processes are active for small energies 
$ |E| \ll |p_yv_F| $. 
For momenta $ p_y $ between these two extrema, 
 also conventional scattering processes are relevant.}
To \rr{avoid} them     
at low energies we must take into 
account (\ref{85}), 
and demand  
that the SL potential $ V(x) $ does not have any  
local minima (maxima) for particles (holes)   
and that the  local minima and maxima are of similar 
absolute potential value. Furthermore we must demand that 
$ V'(x) \approx {\rm const} $ between the local minima and  
maxima. The smooth forms of the symmetric two-step potential 
belong to a class of potentials fulfilling these requirements.

We point  out that these requirements 
\rr{
are not necessary but sufficient to  
determine} the whole low energy region of the lowest energy band 
for a given SL potential within the semiclassical method discussed 
in this paper. 
The reason that these requirements are not necessary lies in the 
fact that the type of scattering depends strongly on the energy of 
the particle or hole. The above requirements hold under the
assumption that $ |E| \ll |p_y v_F| $, and this
does not have to be  
fulfilled for certain momentum values $ p_y $.

\subsection{Constructing SLs  for focusing electron beams}  
In the following, we restrict 
ourselves to unidirectional SLs with a mirror symmetry and an 
additional point symmetry at the origin similar, to the sinus  
potential discussed above. As was shown in the last section, 
this requirement is necessary to find new Dirac points on the $ E=0 $ axis. 
For $ E=0$  we obtain from (\ref{120})
$ S\equiv S_1=-S_2 $ and 
$ \rm{arg}[\overrightarrow{R}] \equiv \rm{arg}[\overrightarrow{R}_1]=
-\rm{arg}[\overrightarrow{R}_2] $, 
that the momentum $ p_y=p_y^n $  of the new Dirac points is  determined by  
\begin{equation} 
 \frac{S}{\hbar} +  \rm{arg}[\overrightarrow{R}] = 
\pi 
(\rr{{\it n}}-1)  \,,   \label{350}  
\end{equation}  
where $ n \in \mathbb{N} $ . 
The number of new Dirac points is then given by 
\begin{equation} 
\left[1 +\frac{1}{\pi} \left(
\frac{S}{\hbar} +  \rm{arg}[\overrightarrow{R}]\right) \right] = n_{\rm max}  \,,   
\label{355}  
\end{equation}  
where we have to  set $ p_y=0 $ in $ S $ and $ \overrightarrow{R} $ entering 
(\ref{355}). 
We have used the abbreviation $ [x] $ as  the larges{\rr t} integer number 
smaller than $ x $.

As mentioned above and can be deduced from 
 (\ref{12}) for the unidirectional step-like SL, 
electrons with a momentum near the central Dirac point are focused strongly
in the direction of the SL wavevector, i.e., $ v_x \gg v_y $, 
especially at potentials where a new Dirac point emerges. It was mentioned 
in Ref.~\onlinecite{Park3} that this phenomenon 
could have technical applications for strong focusing of 
electron beams in graphene. Of course, true focusing of an electron beam   
has the additional requirement that $ v_y=0 $  in the vicinity of  
a specific momentum, and not only exactly for that momentum. 
Such energy dispersions were in fact found in photonic 
crystals \cite{Kosaka1,Rakich1}.   
Within the semiclassical approximation, by solving (\ref{350}) in a nontrivial momentum region, we are now able to 
construct potentials showing exactly 
such a behavior. For doing this we restrict ourselves
to SL potentials in the large $V$-regime of the form  
$ V \sin[2 \pi (x-d/2 -\delta d)/d] $ for $ d/2 \le x_1 \le x \le 3 d/4 $.
Note that due to its symmetry only the discussion of the positive branch 
of the potential, i.e. for $x$-values where 
$ d/2 \le x \le 3 d/4 $, is sufficient.
The value $ x_1 $ is given by the condition  
that (\ref{350}) is fulfilled for the momentum $ p^1_y= V(x_1)/v_F$ where 
we only consider in the following $ n=1 $. 

\begin{figure}
\begin{center}
\includegraphics[clip,height=5.5cm,width=8.5cm]{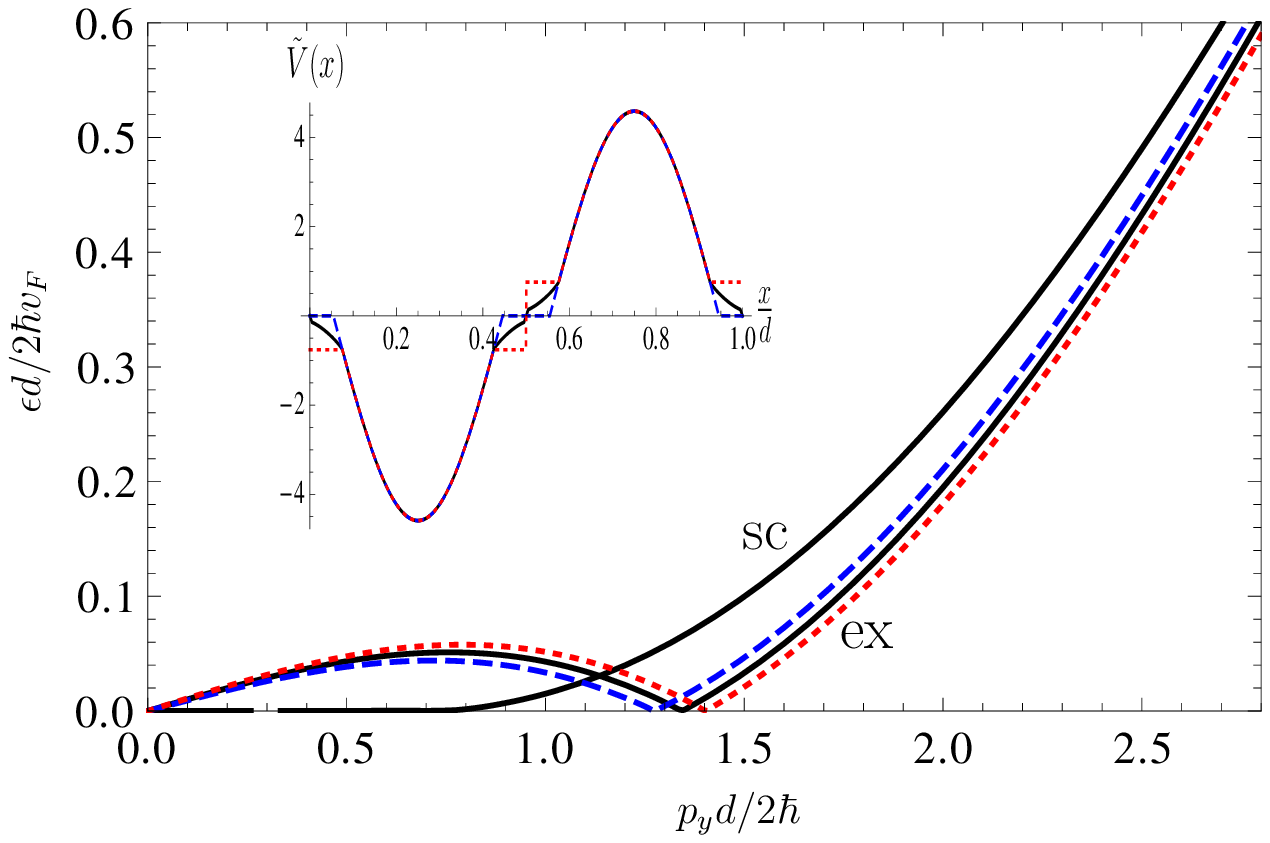}
\end{center}
\vspace{-2em}\caption{The (black) solid curve (sc) is  
the semiclassically calculated energy spectrum for the potential $ V_f $ at 
$ p_x=0 $  
which is shown as the black solid curve in the inset. The other curves (ex) in 
the main panel are the lowest energy spectra calculated by an exact numerical
diagonalization calculation for the potential 
$ V_f $ and its variations shown in the inset.}  
\end{figure} 
 
We now determine the potential $ V_f(x) $ for $ d/2 \le x \le x_1 $ by solving 
(\ref{350}) iteratively. Here we determine $ \delta d $ and $ V $ such  
that $ V_f(d/2) \approx 0 $ and further that the momentum value at the 
first Dirac point $ p^1_y $ is maximal.
With these requirement we obtain 
$ 2 \pi \delta d/d = 0.437 $ and $ \tilde{V}= 4.58 $. 
 
Within the semiclassical approximation this leads to the fact that 
the first side-valley Dirac point $ p_y^1 $ 
and the central Dirac point are   
connected by a flat energy dispersion curve with zero energy. 
We show in the inset in Fig.~6 as the black solid curve 
the potential $ V_f(x) $ obtained in this way. 
The (red) dotted and the (blue) 
dashed  potential curves are variations of $ V_f $ being different   
at $x< x_1 $-values. The black solid curve 
denoted with sc  in the main panel in Fig.~6 shows then the semiclassically 
calculated lowest energy spectrum by using (\ref{120}). Indeed we 
obtain  
a flat energy curve around the central Dirac point. The other 
energy curves shown in the Figure are calculated by using the exact numerical 
diagonalization method. The various exact diagonalization curves 
in Fig.~6 correspond to the potential 
variations shown in the inset. From the (black) solid 
energy curve can be seen that the flatness of the semiclassical 
approximation in fact vanishes within the numerical diagonalization 
calculation. Note that this even holds when going to a high basis 
number in  the exact numerical diagonalization calculation.    
This shows that the semiclassical approximation 
fails here for the constructed potential $ V_f $, at least to the extend 
of having a flat  energy spectrum close to the central  
Dirac point. The reason for this failure presumably comes from the fact 
that at the penetration point $ x_{p21}=x_1 $ the condition 
(\ref{85}) is no longer 
fulfilled. Note that $ V_f $ even gets more shallow when choosing 
larger $x_1 p^1_y $ values where now the energy 
plateau seen in the semiclassical construction cannot be 
extended to $p_y d=0 $.  

From Fig.~6 we even obtain from Fig.~6 that the energy curves of the potential 
variations of $ V_f $ shown in the inset 
do not vary much around the  central Dirac point. 
We consider this as a hint that presumably the whole attempt of finding an  
SL potential with a flat region in the energy spectrum, with one electron and 
one hole region in the fundamental cell,  seems doomed to fail.
Note also that we carried out further numerical calculations  
with variations of the SL potential which turned out to be unsuccessfull
as well.     
It was shown in Ref.~\onlinecite{Sun1} that such a scheme can be successful 
when considering more complicated SLs. 
In that paper it was  shown that a SL with one electron and one hole region  
and an additional small modulation of the potential strengths 
over many fundamental cells of the SL 
can lead to energy spectra with a flat behavior around the Dirac points.        

\begin{figure*}
\begin{center}
\includegraphics[clip,height=8.5cm,width=16.5cm]{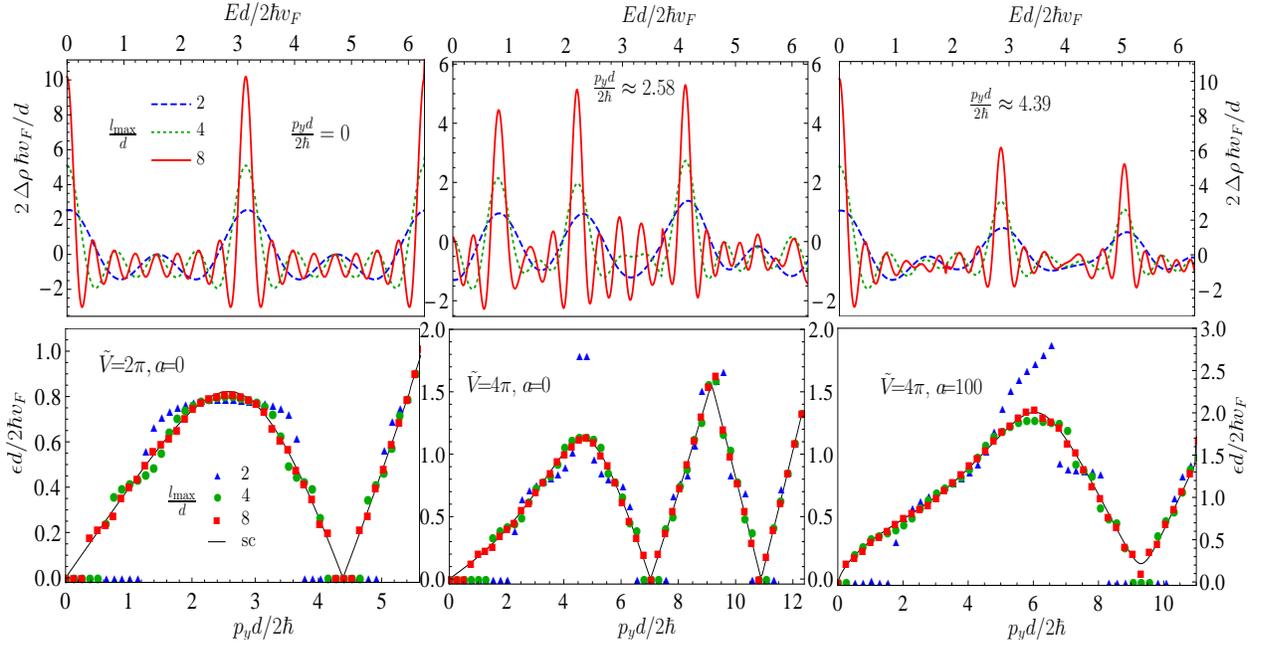}
\end{center}
\vspace{-2em}\caption{Upper panels show the dimensionless density of states 
$ 2 \Delta \hbar v_F \rho/d  $ as a function of the dimensionless energy 
$ E d/ 2 \hbar v_F $ for various maximal orbit lengths 
$ l_{\rm max} $ at $ p_x=0 $. 
The SL potential is given by the non-deformed sinus potential 
(\ref{130}) with $ \tilde{V}=2 \pi $ and $ a=0 $.  
Th density of states is calculated for the central valley 
where $ p_yd/2\hbar =0 $ (left panel), 
at transversal momentum value $ p_yd/2 \hbar =2.58 $
where we found the maximum 
of the lowest energy band (middle 
panel), and at the new side valley Dirac-point 
momentum $ p_yd/2 \hbar =4.39 $ (right panel). 
Lowers panels show the energy of the lowest maximum value of 
the density of states $ \Delta \rho $ for various orbit 
lengths $ l_{\rm max} $. We compare these values  with the 
semi-classical spectrum calculated 
by (\ref{120}) (black curve). We show this   
 for the sinus potential (\ref{130}) with $ \tilde{V}=2 \pi $, $a=0 $ (left panel);  $ \tilde{V}=4 \pi $, $a=0 $ (middle panel)  and 
 $ \tilde{V}=4 \pi $, $a=100 $ (right panel). 
}       
\end{figure*}

\section{Semiclassical density of states} 
The generalization of the above results to general two-dimensional SLs 
via a semiclassical wavefunction solution of Eq.~(\ref{10}) 
is not possible. 
In the case of non-relativistic quantum mechanical systems  this can 
be carried out only for integrable systems  \cite{Brack1}. 
This result is modified in relativistic systems 
mainly due to the existence of the additional phase factor $ \phi(x) $ 
in (\ref{20}) \cite{Littlejohn1, Keppeler1}. 
One way out of this dilemma is by calculating the density of states 
semiclassically with a formalism developed by Gutzwiller \cite{Gutzwiller1}.
The eigenvalue spectrum is then determined from the calculated density 
of states.    

For the unidirectional 
SL with one electron and one hole region  
per fundamental cell we obtain for the density of states 
\cite{Bolte1, Carmier1} 
$ \rho(E)= \overline{\rho}(E)+ \Delta \rho(E) $.   
Here $ \overline{\rho} $ is the average density of states,  
given by 
\begin{equation} 
\overline{\rho}(E) \approx \frac{1}{\pi \hbar} 
\int^d_0 dx {\rm Re}\left[\frac{|E-V(x)|}{ 
v_F^2 \sqrt{\frac{(E-V(x))^2}{v_F^2}-p_y^2}}\right ]\,.  \label{135}   
\end{equation}    
The fluctuating part is given within a semiclassical approximation by 
\begin{equation} 
\Delta \rho(E)\! = \!  \frac{1}{\pi \hbar } {\rm Re}\!  \left[\! \sum_p 
\! T_p \! \sum_{\nu=1}^\infty \! 
[T^{\tau(p)} R^{\sigma(p)}]^{\nu} e^{-i \nu \gamma(p) \frac{p_x d}{\hbar} } 
e^{i \nu \frac{S_p}{\hbar}} \right] \! .    \label{140}  
\end{equation}  
The sum $ p $ in (\ref{140}) 
runs over the primitive periodic orbits of particles $ E-V(x)>0 $ or 
holes  $ E-V(x)<0 $, respectively. Particles  and holes are 
transformed into  each other at the penetration points. 
The configuration space of the orbits is given by the 
fundamental cell of the SL 
with periodic (circular) boundary conditions. $ T_p $ is the required time  
for the particle or  hole for 
passing the primitive orbit. $ T^{\tau(p)} $ stands for 
$ T^{\tau(p)}= 
\overleftarrow{T}_1^{\overleftarrow{\tau}_1(p)} 
\overrightarrow{T}_1^{\overrightarrow{\tau}_1(p)} 
\overleftarrow{T}_2^{\overleftarrow{\tau}_2(p)} 
\overrightarrow{T}_2^{\overrightarrow{\tau}_2(p)} $. 
Here $ \overrightarrow{\tau}_i(p) $ 
$ (\overleftarrow{\tau}_i(p)) $ is the number of 
transmissions from left (right) to right (left) 
through the potential barrier $ i $ in the primitive orbit. $ R^{\sigma(p)} $ 
is the corresponding total reflection coefficient. $ S_p $ is the 
\rr{eikonal} of the primitive orbit, i.e.  
$ S_p=n_1 S_1 + n_2 S_2 $ where $ n_1 $ and 
$ n_2 $ are the number of transitions of the particle regions 
$ E-V(x) >0 $ and   hole regions $ E-V(x) <0 $. 
$ \gamma(p) $  is the winding number of the primitive orbit on the 
circle representing the fundamental cell.

In order to derive Eq.~(\ref{140}), we used the ray splitting 
generalization of Gutzwiller's trace formula first discussed 
in Ref.~\onlinecite{Couchman1}. 
There it was shown that a ray splitting boundary in 
an integrable system can cause additional sign of chaos 
in the energy spectrum. One 
of the simplest systems with ray splitting is that of a  
non-relativistic electron in an infinite one-dimensional 
square well with a discontinuous step inside the well 
\cite{Dabaghian1,Dabaghian2,Timberlake1}. This system, and also  
our system represented by the density of states (\ref{140}) 
can be discussed using the formalism of quantum graphs \cite{Kottos1}. 
With the help of the methods used in this reference
one can directly show the connection between the density of states 
(\ref{140}) and the corresponding energy spectrum represented by equation 
(\ref{120}).

By using that the absolute value of the particle 
velocity is given by $ v_f \sqrt{(E-V(x))^2-p_y^2 v_F^2}/|E-V(x)| $ 
one can easily determine $ T_p $ for every primitive orbit by  
integrating the inverse velocity over the orbit. It is well known 
\cite{Brack1} that a  renumbering of the summands in (\ref{140}) can lead 
to divergent subseries. A well-behaved approximation should be achieved 
by sorting the terms in (\ref{140}) with respect to their maximal 
orbit length $ l_{\rm max} $. This means that for 
$ l_{\rm max} =2 m d $ with $ m \in \mathbb{N} $ we have to take 
into account in  (\ref{140}) 
all orbits with lengths less than or equal to $ 2m d $.  
In the upper row in Fig.~7  we show  
$ \Delta \rho $ for $ l_{\rm max}= 2d$ , $4 d $ and $ 8d $ 
for the sinus potential (\ref{130}) with $ \tilde{V}= 2 \pi $,  
$ a=0 $ and $ p_x=0 $. These panels are calculated for $ p_y d/\hbar $ 
values where the lowest energy band  (cf. Fig.~4) has its two 
Dirac points 
(left and right panel) and further where the band 
has its  local maximum (mid panel).
By comparing the curves with the corresponding energy spectrum in 
Fig.~4 we obtain that the energy values of the lowest energy band 
correspond to the smallest energy maximum in 
$ \Delta \rho $. This happens even when we take into account  
only small $ l_{\rm max} $ orbit lengths. We note that the higher energy maxima 
of $ \Delta \rho $  in Fig.~7  correspond to higher energy bands. 
Next we try to reproduce 
the lowest energy band for various (deformed) sinus potentials and 
$ l_{\rm max}= 2d $, $ 4 d $ and $ 8d $ from the lowest energy maximum in 
$ \Delta \rho $. We compare our result in Fig.~7 
with the semiclassically calculated 
lowest energy band by using (\ref{120}) ((black) straight curves). 
We obtain from the figure that  
the new Dirac points even show up for small $ l_{\rm max} $ in form 
of a plateau 
at zero energy where its extension is rapidly decreasing for higher 
$ l_{\rm max} $-values. Note that the maximum criterium used here for 
determing the spectrum from $ \Delta \rho $ is different  from the common 
approaches used for determing the full  energy spectrum for 
systems  in the field of quantum chaos. 
There commonly the condition that the  
integration of the full density of states between 
two non-degenerate energy levels should give 
the value one is used.  Since we are only interested in the lowest 
energy level and furthermore  the lowest energy band and the first excited 
energy band are well separated, such an approach is not necessary here.

\section{Conductivities} 
Next, we calculate the conductivities parallel and orthogonal to 
the SL wavevector by using the semiclassical wavefunction 
(\ref{20}) and energy dispersion (\ref{120}) for SLs with 
a point symmetry at zero energy for half-filling. 
We thereby restrict ourselves  to the ballistic transport regime. 
Note that ballistic transport was seen for graphene samples without a SL 
up to the micron length \cite{Morozov1,Du1, Bolotin1}. 
Taking into account also the small interlattice spacing of $ 1.4 $~\AA~  
in graphene makes the ballistic transport regime relevant even for large 
superlattices.

There are various techniques in the literature for calculating ballistic 
conductivities in graphene. Below, we will use a formalism firstly    
introduced in  Ref.~\onlinecite{Lewkowicz2} for graphene without a SL. 
In this approach the linear ballistic transport is calculated 
as a response to an electric field given by a temporal gauge field of 
the form $ {\bf A}=- c {\bf E} t \Theta(t) $, where $ {\bf E} $ is the external 
electric field.  

There are 
also Kubo-like formalisms in the literature using gauge fields of the spatial 
form. These have the disadvantage that the calculated 
conductivities in these formalisms are only well defined  
up to a numerical prefactor which depends on the order of taking 
the zero-temperature, zero-frequency, and zero-damping limit 
\cite{Ziegler1, Ryu1}. The simplest versions of both of these formalisms 
above work for non-doped leads. 
For heavily doped leads a Landauer-like transfer matrix formalism 
\cite{Katsnelson1,Tworzydlo1} can be found in the literature for 
SL-free pristine graphene.  
Here evanescent modes give the dominant contribution to the conductivity. 
These modes do not longer fulfill the Bloch condition which makes 
it complicated to find analytical conductivity results for general smooth SLs. 
A further complication comes from the fact that in using 
the semiclassical approach one has to demand that the leads are coupled 
to the graphene system in a smooth way introducing a new parameter to the 
system. Finally we note the important fact that the Landauer formalism 
for heavily doped leads and the temporal gauge formalism for non-doped leads, 
which we will use below, result in numerical similar conductivity 
values for pristine graphene.

The lowest band eigenvalue spectrum is given by (\ref{120}) which was 
effectively calculated from the eigenvalues of the matrix 
$ \underline{A} $.
By using (\ref{120}) we obtain the following energy dispersion around the 
Dirac points, i.e. for $ |\epsilon_s| d I(-1,0)/2 v_F S \ll 1 $,  
\mg{{OK? else rewrite formula understandably}}
\begin{align} \!\!\!\!\!\!\!\!\!
& \!\!\!\!\!\! \epsilon_s  =  s \frac{2 \hbar v_F}{d}\frac{1}{I(-1,0)}\Bigg[ |T|^2
  \sin^2\left(\frac{p_x d}{2 \hbar}\right)+ \label{510} \\
&
\!+|R|^2
\sin^2\left( \frac{S+ {\rm arg}[\overrightarrow{R}]} { \hbar} \right)+ 
\frac{1}{2}(1- |R|^2-|T|^2)\Bigg]^{1/2}  \nonumber 
 \end{align}
where $ s= 1 $ for the conduction band and $ s=-1 $ 
for the valence band.  Here we use the abbreviation 
$ {\rm arg}[ \overrightarrow{R}] \equiv 
({\rm arg} [\overrightarrow{R}_1] -{\rm arg}[ 
\overrightarrow{R}_2])/2 $ and 
$ |R| \equiv  \sqrt{|\overrightarrow{R}_1||\overrightarrow{R}_2|} $,  
$ |T| \equiv  \sqrt{|\overrightarrow{T}_1||\overrightarrow{T}_2|} $ 
and denote  
$ \overrightarrow{R} \equiv |R| e^{i {\rm arg}[\overrightarrow{R}]} $
for energies $ E=0 $. 
The function $ I(n_1,n_2) $ is defined by 
\begin{equation} 
I(n_1,n_2)=\frac{2}{d}  \int_{x_{pi1}}^{x_{pi2}} \! \! \!  dx \, 
  \frac{\sqrt{
\left(\frac{V(x)}{v_F}\right)^2 -p_y^2}^{n_1}  p^{n_2}_y}{|V(x)/v_F|^{n_1+n_2}}
\,. 
\label{515}    
\end{equation}

In the following, we will use the eigenfunctions of the matrix
$ \underline{A} $ which was defined following Eq.~(\ref{120}). 
These are given in the vicinity 
of the Dirac points, i.e. for   
$ |\epsilon_s| d I(-1,0)/2 v_F S \ll 1 $,  by  
\begin{equation} 
 {\bf EV} \! \! \approx \!   
 i \! \left( \! \! { \sin\! \left[\! \frac{Ed}{\hbar v_F}  I(-1,0)\! \right] 
 \! \! - \! \! \sin\left[2 \frac{S}{\hbar}\! \!  + \! \! 2 
{\rm arg}[\overrightarrow{R}]\right] |R|^2 \! \! + \! \!  
  \frac{p_x d}{\hbar} |T|^2  \atop 
  -  e^{ i \frac{S}{\hbar} } \left( |R_1| e^{ i \frac{S}{\hbar} 
+ {\rm arg}[\overrightarrow{R}_1]} -
|R_2| e^{ -i \frac{S}{\hbar} + {\rm arg}[\overrightarrow{R}_2]} \! \right)    \! 
} \right)
  \label{520}
 \end{equation}

The lowest\rr{-}band eigenfunctions for electrons in the SL are  then 
given for $ 0 \le x \le d $ by 
\begin{align} 
& {\bf u}_{s}(x)\approx 
\bigg( \underline{M}_1\underline{N}_1(x) 
\Theta(x-x_{p11})\Theta(x_{p12}-x)  \nonumber \\ 
& + \underline{N}_2(x)
\underline{M}_1\underline{N}_1(x_{p12}) 
\Theta(x-x_{p21})\Theta(x_{p22}-x)\bigg) {\bf EV} /N.     \label{530}  
\end{align}  
Here $ N $ is a normalization constant. 
Note that we \rr{omitted} here once more semiclassical phase factors (\ref{40})
as previously in (\ref{120}).    
We show below that they will in fact not contribute to the conductivity
within the semiclassical approximation. 
Furthermore we idealized in (\ref{530}) the whole wavefunction by setting  
it to zero in the classical forbidden region. We 
will also justify this assumption below.

Next we calculate the dc-response in the SL system. 
This is done in the gauge  
$ {\bf A}=- c {\bf E} t \Theta(t) $. 
The conductivity in the $i$-th direction in the lowest energy level 
approximation valid for $ t \to \infty $  
is then given by \cite{Lewkowicz2,Dietel2} 
\begin{equation} 
   \tilde{\sigma}_{ii}\! = \! \frac{-4 e v_F}{(2 \pi)^2}
 \int_{\rm BZ}  \! \! \! \frac{d^2p}{\hbar^2}  
\mbox{Re}[ e^{-\frac{i}{\hbar} 
\Delta \epsilon t} \langle {\bf u}_{-1}|  
\sigma_{i}| {\bf u}_{+1} \rangle \xi_+(t) ]    \label{575}     ,
\end{equation} 
with 
\begin{equation}   
\xi_+(t)=  -i \frac{ e v_F}{\hbar}  
\! \! \int_{t'=0}^t \! \! \! \! \! dt' \! \! 
\int_{t''=-\infty}^{t'} \! \! \! \! \! \! \! dt'' 
{\cal T}(t'')    ,      \label{585}
\end{equation} 
and the transition matrix element $ {\cal T}=
e^{\frac{i}{\hbar} \Delta \epsilon t} 
\langle {\bf u}_1| \sigma_{i} |{\bf u}_{-1} \rangle$. 
The value $ \Delta \epsilon$ 
is given by the energy gap $ \Delta \epsilon= \epsilon_1- 
\epsilon_{-1} $ for an electron with momentum $ p_y $. 
The integral in (\ref{575}) is carried out over the full 
Brillouin zone.

In the following
\rr{we separately calculate 
the contribution of 
every energy valley
 to the 
momentum integral in (\ref{575}), i.e.,}
\begin{equation}  
 \tilde{\sigma}_{ii}= \sum_{n=0} \tilde{\sigma}^n_{ii}
(2- \delta_{n,0}) .   \label{593} 
\end{equation} 
For large times one can restrict  
the $ p_y$-integrals of Eq.~(\ref{575}) 
to the vicinity of the valley center $p_y^n $ 
in $ {\cal T} $ where $ p_y^n $ 
is determined by (\ref{350}) for 
$ n > 0 $ and $ p^{0}_y=0 $ for the central valley.
The factor two in (\ref{593}) \rr{takes into account the mirror symmetry 
of the energy spectrum with respect to $p_ y $, 
such that we may consider only
$ p^{n}_y \geq 0 $ in (\ref{593}).     }

For calculating the conductivity $ \tilde{\sigma}^n_{ii} $, we first 
have to \rr{determine} the matrix element 
$ \langle {\bf u}_1| \sigma_{i} |{\bf u}_{-1} \rangle $. 
We \rr{apply
the semiclassical approximation by assuming 
$ \hbar  $ being small enough to neglect integrals of the 
form  $ \int dx e^{i 2 \int^x dx'S_i(x')/\hbar} $ in comparison to the integrals 
$ \int dx e^{i 0  \int^x dx'S_i(x')/\hbar}$}.\mg{check exponents 2 and 
i0 in marked left column} 
\mn{check exponents 2 and i0} \rr{On similar grounds we
may also neglect
the matrix contributions in the classical forbidden regions.}  
 From this argument it becomes   
evident that the semiclassical phases $ \phi(x) $ (\ref{40}) 
will not contribute to 
$ \langle  {\bf u}_1| \sigma_{i} |{\bf u}_{-1} \rangle $ since 
the integrand in (\ref{40}) is inverse proportional to $ 1/p_x(x') $.

By using (\ref{530})\rr{--}(\ref{585}) we obtain the following 
conductivities  
\begin{equation} 
\tilde{\sigma}^n_{ii}=\frac{1}{2} 
\frac{e^2}{h}  \frac{O_i}{|v_x v_y|/v_F^2}         \label{595} 
\end{equation} 
with $ v_x $ and $ v_y $ being the electron velocities at the Dirac point. 
By using (\ref{510}) we obtain
\begin{align} 
& |v_x| = |T| \frac{v_F}{I(-1,0)}\,,         \label{600}  \\
& |v_y|= \frac{2 v_F \hbar }{I(-1,0) d} \\
&\times\partial_{p_y}\sqrt{|R|^2 
\sin^2\left( \frac{S}{ \hbar}+ {\rm arg}[\overrightarrow{R}] \right)
 \rr{+  
\frac{1}{2}(1- |R|^2-|T|^2)}}\,. \nonumber 
\end{align}  
The absolute square of the transition matrix elements are given by 
\begin{align} 
&  O_x = \int_0^{2 \pi}  d \vartheta \frac{(N^d_x F^{1}_{-})^2}{
|N^d F^2_{+}+ N^{nd} F^3_{+}||N^d F^2_{-}+ N^{nd} F^3_{-}|}, \nonumber 
 \\ 
& O_y = \int_0^{2 \pi}  d \vartheta \frac{(N^d_y F^{1}_{+}+N^{nd}_y 
F^{4})^2}{
|N^d F^2_{+}+ N^{nd} F^3_{+}||N^d F^2_{-}+ N^{nd} F^3_{-}|} \label{605} 
\end{align} 
with 
\begin{align} 
&  N^d =  \frac{2}{T^2}\bigg\{\left(\frac{5}{2}+2 |R|^2\right)I(-1,0)  \nonumber \\
& + 
{\rm Re}[\overrightarrow{R}][I(-1,2)-I(-1,0)]+ 2 {\rm Im}[\overrightarrow{R}] 
I(0,1)\bigg\} \,, \nonumber  \\
&  N^{nd} =  \frac{4}{|T|^2}|R| I(-1,0) \,, \nonumber \\
& N^d_x = 4 I(0,0) \,,  \nonumber  \\
&  N^d_y =  \frac{4}{|T|^2}\bigg\{
{\rm Re}[\overrightarrow{R}][I(-1,1)+{\rm Im}[\overrightarrow{R}]I(0,0)\bigg\} \,,  \nonumber \\
& N^{nd}_y = \frac{4}{|T|^2} |R| I(-1,1) \label{610},
\end{align}    
and 
\begin{align}
& F^{1}_{\pm} \!  = \! 
H_+(\vartheta) H_-(\vartheta)\!  \pm \!   \cos^2(\vartheta)\,  , \,  
 F^2_{\pm} \!  = \!   H^2_{\pm} (\vartheta) \! + \! \cos^2(\vartheta),   \nonumber \\
& F^3_{\pm} \!  = \!   2 \cos(\vartheta) H_{\pm}(\vartheta) \,, \, 
F^4 \!  = \!   - 2 \cos(\vartheta) (H_{\pm}(\vartheta)\mp 1 ) 
  \label{620},
\end{align} 
where  
\begin{equation}  
H_\pm(\vartheta) = \pm 1 - |R| \cos(\vartheta)+|T| \sin(\vartheta). \label{625} 
\end{equation}  

In the following, we further specify the parameters in (\ref{595}). 
For the side-valleys $ n >0 $, the value $ p_y $ in the expressions 
(\ref{595})-(\ref{620}) is given by the side-valley 
Dirac point momentum determined by (\ref{350}).      
In this case we obtain for $ v_y $ (\ref{600}) 
\begin{equation} 
|v_y|= \frac{2 v_F }{I(-1,0)}\left[ -\frac{1}{2} I(-1,1) + 
\frac{\hbar}{d} \partial_{p_y} {\rm arg}[\overrightarrow{R}] \right] |R|   \label{630},
\end{equation} 
with 
\begin{align} 
& \frac{\hbar}{d} \partial_{p_y} {\rm arg}[\overrightarrow{R}] =\frac{1}{4}  
[{\rm Re}[\Psi(i \tilde{p}_y^2)]- \log(\tilde{p}_y^2)]
\bigg[\left(\frac{x_{p21}-x_{p12}}{d} \right)   \nonumber \\      
& + 
\frac{p_y v_F }{d}\left(\frac{1}{V'(x_{p21})}+\frac{1}{V'(x_{p12})}\right) 
\bigg],  \label{635}    
\end{align} 
where $ \Psi $ is the digamma function. 
   
For the central valley $ n=0 $  we have 
\begin{equation} 
\frac{\hbar}{d} \partial_{p_y} |R_i|= 
\sqrt{\frac{\pi \hbar v_F}{d^2 V'(x_{pi})}}\,. \label{650}
\end{equation} 
Here $ x_{pi} $ is the \rr{$i$}-th intersection point of the SL potential and 
the x-axis, i.e. $ x_{p1}=d/2 $ and $ x_{p2}=0 $. 
The $ p_y $ momentum value in the 
expressions (\ref{595})-(\ref{635}) is then given by 
$ p_y=0 $.
The electron velocity in $ y$-direction is for $n=0 $  given by 
\begin{align} 
&\!\!\!\!|v_y  |\!=\!\frac{2 \hbar v_F }{d I(-1,0)}\times\Bigg[
\sin^2\left(\frac{S}{\hbar} -\frac{3 \pi}{4}\right) 
(\partial_{p_y} |R_1|)
 (\partial_{p_y} |R_2|)      \nonumber \\
& ~~~~~ ~~~~~ ~~~~~\quad\quad\quad
+
\rr{\frac{1}{4} 
\left(\partial_{p_y} |R_1|-
\partial_{p_y} |R_2|\right)^2\Bigg]^{1/2}\!\!.}       \label{660} 
\end{align} 

The only non-zero 
values in (\ref{610}) for $ p_y=0 $ are given by $ N^d= 5 $ and $ N_x^d= 4 $.
This leads to  $ O_y =  0 $ and $ O_x  = 16\pi/25 $.  
       
For a step-like SL potential $ V(x) = V \chi(x) $, 
the dc-conductivities are given by \cite{Burset1,Dietel2}   
\begin{equation} 
  \tilde{\sigma}^n_{xx} =  \frac{e^2}{h} \, \frac{\pi}{2}
\tilde{\alpha}^2_0   \frac{1}{|\Gamma_n|}  
\quad, \quad  \tilde{\sigma}^n_{yy} = \frac{e^2}{h} \frac{\pi}{2}
\frac{1}{\tilde{\alpha}^2_0} |\Gamma_n|\,.    \label{700}     
\end{equation}
with $ \tilde{\alpha}_0= \pi n / \tilde{V} $,  $ \Gamma_n = 
(\tilde{V}^{2}-(\pi n)^{2})/\tilde{V}^{2} $, $\tilde{V}= V d/\hbar v_F 2$. 
The index $ n $ denotes  
the valleys $ n=1,\ldots,[\tilde{V}/\pi] $, 
where $ [x] $ is the largest integer value smaller than $ x $. 
Here $ n=1 $ denotes the outermost 
valley, and $ n=[\tilde{V}/\pi] $  the first valley next to  
the central one.     
For the central valley, we have $ \Gamma_0 = \sin(\tilde{V})/\tilde{V} $ and 
$ \tilde{\alpha}_0=1 $.

We show in Fig.~8 the conductivities $ \tilde{\sigma}^n_{xx} $ 
($ \tilde{\sigma}^n_{yy}$)  in the left (right) panel as a function of 
the potential strength $ \tilde{V} $ for the non-deformed sinus potential 
(\ref{130}) with $ a=0 $. We deduce from the figure that for 
$ \tilde{\sigma}_{xx} $ the central valley contribution 
$ \tilde{\sigma}^0_{xx} $ 
to the conductivity is most relevant where for 
$ \tilde{\sigma}_{yy} $ the outermost  
valley $ \tilde{\sigma}^{1}_{yy} $  contributes the most. 
This is in accordance with the case of the step-like potential 
$ V(x)= V \chi(x) $ (\ref{700}). We can even infer from the figure that in 
practice one can neglect the non-dominant valleys in expression  
(\ref{593}). This is in contrast to the step-like case where the 
non-dominant valley contributions are much larger. 
The reason lies in the fact that 
for smooth potentials $ V(x) $, $ \tilde{\sigma}^n_{ii} $ (\ref{595}) 
contains exponential damping terms as a function of $ p_y $ 
via their dependence on the transmission coefficient 
$ |T| $. We obtain from (\ref{595}) 
$ \tilde{\sigma}^n_{xx} \sim  |T|^3 $, $ \tilde{\sigma}^n_{yy} \sim 1/|T| $.
The exponentially  vanishing behavior of the transmission coefficient 
$ |T| $ for large $ \tilde{p}_y^2 $ in smooth 
potentials is caused by the exponential damping of the wavefunction 
in the classically forbidden region. In contrast to this, the 
transmission coefficient $ |T| $ for a step-like potential $ V(x) $ is  
decreasing algebraically as a function of  $ \tilde{p}^2_y $.

One can understand the $ |T| $-behavior of (\ref{595}) also in the 
following heuristic way. The finite quantum conductivity in pristine 
graphene 
is heuristically  conceived by taking into account Einstein's law for 
classical diffusive scattering. 
There the  conductivity is proportional to the density of 
states multiplied by the diffusion constant. As in every two dimensional system 
for infinite small scattering the effective 
diffusion constant is infinite.  At the same time, in contrast to 
two-dimensional metals where the density of states is constant,  
it vanishes in graphene at the Dirac point, 
leaving the  total conductivity as a constant.
By the application of a SL in \rr{$x$-direction,} the diffusion in y-direction 
is in first approximation the same 
in pristine graphene, but the density of states 
scales with $ 1/ |T| $ (\ref{510}), leading to 
$ \tilde{\sigma}_{yy}\sim 1/|T| $. 
In contrast to this,  \rr{
the scattering in the  $x$-direction
for graphene with a superimposed SL
is for $ |T| \ll 1 $  
mainly diffusive}, with a diffusion constant 
$ \sim |T|^2 $. The density of states still scales with 
$ 1/|T| $. Since the density of states vanishes at the Dirac point 
we obtain an extra $ |T|^2 $-term in $ \tilde{\sigma}_{xx} $, leading to 
$ \tilde{\sigma}_{xx} \sim |T|^3 $. More precisely this extra $ |T|^2$ term 
follows from the averaging of the density of states over the inverse 
coherence time of the wavefunctions $ \sim |T|^2 $ in Einstein's law.
From this argument it is even easier to  understand the finite conductivity  
of the SL-free pristine graphene system in the limit of 
infinite small scattering.

\begin{figure}
\begin{center}
\includegraphics[clip,height=4.5cm,width=8.5cm]{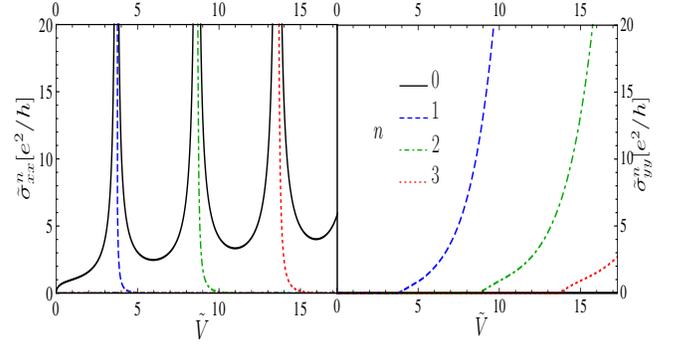}
\vspace{-2em}
\end{center}
\caption{Left panel shows the valley contribution $ \tilde{\sigma}_{xx}^n $ 
to the conductivity in parallel direction to the SL wavevector 
calculated within the semiclassical approximation (\ref{595}) as a function 
of the potential strength.  
Here we used the non-deformed sinus potential (\ref{130}) for $ a=0 $ 
as the SL. 
Right panel shows $ \tilde{\sigma}^n_{yy} $ for the same potentials.}  
\end{figure} 

From (\ref{595}) we deduce that even 
for small SLs where only the central Dirac point is present, 
$ \tilde{\sigma}^0_{yy} $ is zero. This is not true for the 
orthogonal conductivity $ \tilde{\sigma}^0_{yy} $ of the step-like
SL system (\ref{700}). In Ref.~\onlinecite{Brey1} the conductivity 
$ \tilde{\sigma}_{xx} $ for the non-deformed finite length sinus SL potential 
(\ref{130}) as a function of $ \tilde{V} $ was 
calculated by using a transfer  matrix method for 
heavily doped graphene leads \cite{Katsnelson1, Tworzydlo1}. 
In the left panel of Fig.~8 we see a good 
quantitative accordance of our result with their curves. 
We consider this as a justification of 
the semiclassical approximation method considered in this paper.        
 
\begin{figure}
\begin{center}
\includegraphics[clip,height=5.5cm,width=8.5cm]{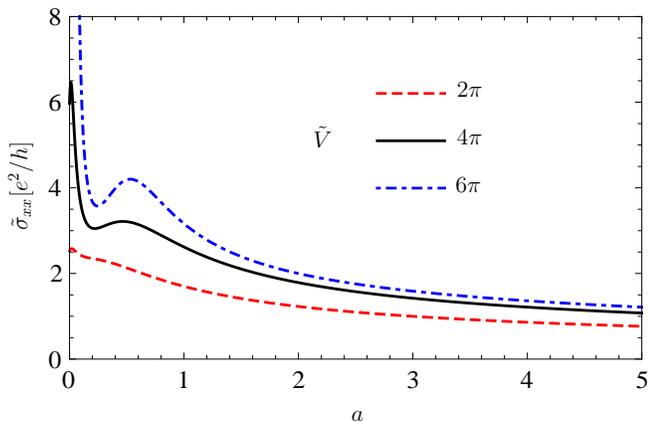}
\end{center}
\vspace{-2em}\caption{\rr{Conductivity} $ \tilde{\sigma}_{xx} $ for a deformed sinus potential 
of the form (\ref{130}) as a function of the deformation 
parameter $ a $ for various potential strengths $ \tilde{V}$.}     
\end{figure}

Finally in Fig.~9, we show $ \tilde{\sigma}_{xx} $ for the deformed 
sinus potentials (\ref{130}) as a function of the deformation parameter 
$ a $ for various potential strengths $ \tilde{V} $. We argued in 
 Sect.~II that in this case only  the central Dirac point exists 
leading to the fact that $ \tilde{\sigma}_{yy}=0  $ within the semiclassical 
approximation. From Fig.~9, we obtain 
local maxima in $ \tilde{\sigma}_{xx} $ at 
certain deformation values $ a $. As can be seen from (\ref{595}) 
with (\ref{660}), these deformation parameters are in a regime  
where (\ref{350}) is fulfilled for values of 
$ n \in \mathbb{N} $ and $ p_y=0 $ with  
$ {\rm arg}[\overrightarrow{R}]=-3 \pi/4 $.

\section{Success, Failures and possible Applications 
of the semiclassical approach}

Just recently, an extensive analysis 
of the semiclassical transmission coefficients of np and npn junctions 
with a comparison to a numerical multistep calculation 
was carried out \cite{Tudorovskiy1}. 
Up to small deviations for small incident angles of the particles, 
the authors find quite good agreement of the semiclassical 
results with their numerics. This is in accordance to the 
good results we found for the energy spectrum  
of the (deformed) sinus SL potentials in Sect. II. By taking  
into account also the good conductivity behavior of the semiclassical 
approximation described in the last section we could conceive the 
following application. 

As already argued in the introduction of this paper, 
due to their spectral and conductive properties,  
electrons in graphene with an overlying SL are interesting systems promising
many applications, which could open new routes to building electronic devices.  
\rr{The ability to construct SL potentials which show this behavior 
for an
energy band with desired conduction properties can be very useful.
We have shown that this is in principle possible within the 
semiclassical approximation in our example in 
Sect.~IIB}. There we reached the goal to construct 
 SL potentials showing a plateau in the energy 
spectrum as a function of the transversal momentum using the
semiclassical approximation. 
Unfortunately, this behavior did not survive when calculating 
the energy spectrum of the constructed SL potential 
with an exact numerical diagonalization method.
The reason lies in the fact that the required smooth 
behavior of the constructed potential, which is necessary 
for the validity of the semiclassical approximation was not given. 
The lesson to be learned  from this example is that  
semiclassically constructed potentials should be further 
crosschecked by additional means,  as e.g. using numerical
methods, in order to be trusted.

\section{Summary} 
\rr{We have analysed}
the behavior of electrons in electrical superlattice potentials
within a semiclassical approximation. We found
this description to work well for smooth superlattice potentials. 
We started in Sect. IIA by introducing    
the semiclassical wave function representation of the quasi-relativistic 
Dirac equation \rr{of} electrons in graphene superimposed by an SL.  
We have \rr{derived transmission and reflection} 
coefficients for Klein tunneling  
through a classical forbidden region, in which a \rr{particle  state is 
converted to a hole  state or vice versa}. Within a 
generalized Bohr-Sommerfeld formalism,
\rr{we have
derived the eigenvalue 
equations for the lowest energy band of a SL with one electron and one hole 
region}
in the fundamental cell, showing only Klein scattering.   
For electrons 
in a SL of a (deformed) sinus shape, we obtain very 
good accordance of the semiclassical energy spectrum 
with the spectrum obtained by exact numerical diagonalization. 
Then we tried  to construct in Sect.~IIB 
potentials having an energy plateau at zero energy, and uncovered  its 
failure when comparing the semiclassical 
energy spectrum of the potential with the exact 
diagonalization method as  already 
described in the last section.
In order to  pave the path to take into account SLs which are not 
unidirectional we calculated in Sect.~III the semiclassical density 
of states within the generalized Gutzwiller trace formula by taking 
into account the beam-splitting extension. Even by considering only 
small length orbits we could reconstruct the energy spectrum
of the lowest band from the density of states maxima. 
This was carried out explicitly for the (deformed) sinus potential SLs.
 
Finally we have calculated in Sect.~IV longitudinal ballistic conductivities 
along and transverse to the SL wavevector 
within the semiclassical 
approximation. Here we have restricted  ourselves  again 
to the simplest point symmetric SLs with one electron and one hole region 
in the fundamental cell where only Klein scattering is important. 
We obtain a good quantitative 
accordance with conductivity curves found in the literature  
for sinus potentials as a function of the potential strength. 
In these calculations a transfer matrix method was used in order to     
calculate the conductivity parallel to the SL wavevector.  
Furthermore we 
obtain, as was formerly shown also for step-like SLs,  
that the conductivity along the 
wavevector of the SL is mainly governed by electrons in the central valley  
whereas the orthogonal 
conductivity is determined mostly  by the conductivity contribution of the outermost 
valley. The contribution of electrons in the central valley 
is zero in the latter case.  
In contrast to the step-like SLs, the neglect of the other non-dominant 
valleys is exponential damped in both cases. 
This is connected to the fact that the 
transmission coefficients for Klein tunneling in smooth potentials  
in contrast to step-like SLs are exponentially small as a function of the 
length of the classically forbidden region and transversal momentum.     

\acknowledgements
The authors acknowledge the support provided by Deutsche Forschungsgemeinschaft
under grant KL 256/42-2.

\end{document}